\documentclass[twocolumn,twoside,slac_two]{revtex4}
\usepackage{graphicx}
\usepackage{fancyhdr}
\usepackage{graphics}
\usepackage{epstopdf}
\usepackage{textpos}
\newcommand{\aerr}[2]{{\:}^{+{\:}#1}_{-{\:}#2}}%

\begin{document}

\title{\centering Heavy Flavour Results at the LHC}


\author{
\centering
\includegraphics[width=0.15\textwidth]{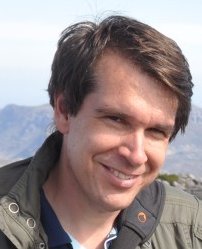} \\
\begin{center}
P. Koppenburg
\end{center}}
\affiliation{\centering Nikhef, Amsterdam, The Netherlands 
\vskip 0.5em On behalf of the LHCb collaboration, including 
Atlas, CMS and Alice results.
}
\begin{abstract}
We present a brief overview of the first flavour physics results at the LHC.
Cross-section for charm and beauty production have been measured by several experiments
and the first competitive results on $D$ and $B$ decays are presented.
\end{abstract}

\maketitle
\thispagestyle{fancy}


\section{Introduction}
Precise measurements of CP violation and searches for rare decays have a high potential
for discovering effects of New Physics. They are a specific task of the LHCb
experiment and are complementary to direct searches performed by general purpose
experiments. CP violation and rare decays are sensitive to new particles and couplings
in an {\it indirect} way via interferences with Standard Model (SM) processes. 
This not only probes a potentially higher mass scale than direct searches for new particles, 
but also gives access to amplitudes and phases of the new couplings.
The $b$ and $c$ quark decays are the best laboratory 
for this programme. 
\section{The LHC as a the new Flavour Factory}
After a very successful decade dominated by the $B$ factories, Belle and Babar,
the LHC is taking over as the new flavour factory. While PEP-2 and KEK-B have produced
around $10^9$ $b\bar b$ pairs during their lifetime, the LHC has produced 
close to $4\cdot10^{12}$ just in 2011 thanks to the very large $b$ cross-section
in high energy proton collisions. Of course the major challenge at the LHC is 
to efficiently collect the most interesting among all these events.
\subsection{The LHCb Experiment}
The LHCb experiment~\cite{Alves:2008zz} is dedicated to precision measurements
of CP violation and rare decays in beauty and charm decays.
Its forward geometry covering the range $2<\eta<5$ 
exploits the dominant heavy flavour production mechanism
at the LHC and covers about 40\% of the differential cross-section.

Among the features unique to LHCb are its high precision vertex detector, which is 
retracted away from the beam at injection and moved as close as $8\:$mm from the beam
during data taking. It is followed by a tracking system located in a dipole
magnet of which the polarity can be reversed, allowing to cancel detector asymmetries
in CP-violation measurements.
A system of two Ring Imaging Cherenkov detectors (RICH) allows
a very good pion/kaon/proton separation over the momentum range $1$--$100\:{\rm GeV}/c$.

Due to the very large total cross section an effective
on-line event selection is required where the rate is reduced by a first level hardware
trigger and further by two levels of software 
triggers~\cite{vanHerwijnen:2010zz}. LHCb uses hadrons, muons, electrons and photons
throughout the trigger chain, thus maximising the trigger efficiency
on all heavy quark decays.

In order not to saturate the trigger and keep the event multiplicity low, the LHCb
experiment is not operating at the maximum LHC luminosity. During most of 2011 LHCb has kept 
their luminosity at the constant value of $3.5\cdot10^{32}\:\rm cm^{-2}s^{-1}$
by displacing the proton beams laterally in real time.
\subsection{Flavour Physics at Atlas, CMS and Alice}
The Atlas~\cite{Aad:2008zzm} and CMS~\cite{Adolphi:2008zzk} detectors are
multi-purpose central detectors optimised for searches of heavy objects. 
At high luminosity, their potential for flavour physics is limited 
by their triggering capabilities and focus mainly on $b$ and charmonium decays involving dimuons.
But at the lower luminosities at which the LHC operated 
during 2010 and part of 2011, a more open trigger
allowed an interesting favour physics programme that is complementary to LHCb's in 
particular for cross-section measurements in the central region.

The Alice~\cite{Aamodt:2008zz} detector is optimised for heavy ion collisions
and, while covering mostly the central region, has similar particle ID and tracking
capabilities as LHCb. They are thus a key player for cross-section measurements
(which are often a necessary normalisation or their QGP programme) but are not
competitive for CP violation and rare decays searches due to the lower luminosity 
at which they operate.

\subsection{Data Samples}
In 2011, Atlas and CMS have been delivered around $5.7\:\rm fb^{-1}$ each, LHCb
$1.2\:\rm fb^{-1}$ and Alice $5\:\rm pb^{-1}$, of which more than 90\% have been recorded
and are useful for physics. The measurements reported below 
use only a fraction of these data, in most cases the $1.1\:\rm fb^{-1}$ (Atlas, CMS) and 
$370\:\rm pb^{-1}$ (LHCb) collected until end of June 2011. Most cross-section measurements
use the lower-luminosity data sample of about $40\:\rm pb^{-1}$ collected during 2010.
\section{Heavy Flavour Production}
The four LHC experiments offer a vast coverage of rapidity:
Atlas and CMS cover the central region up to $2.5$, Alice $0$--$1$ and $2.5$--$4$ and
LHCb $2$--$5.5$. A combination of differential cross-section measurements 
would allow to cover the range $0<|\eta|<5.5$, but in most cases such combinations 
have not yet been performed. Yet, many measurements from the various experiments are
available and give a good picture of heavy flavour production in $pp$ collisions
at $\sqrt{s}=7\:\rm TeV$.
\subsection{Charmonium}
\begin{figure}
\includegraphics[width=\columnwidth]{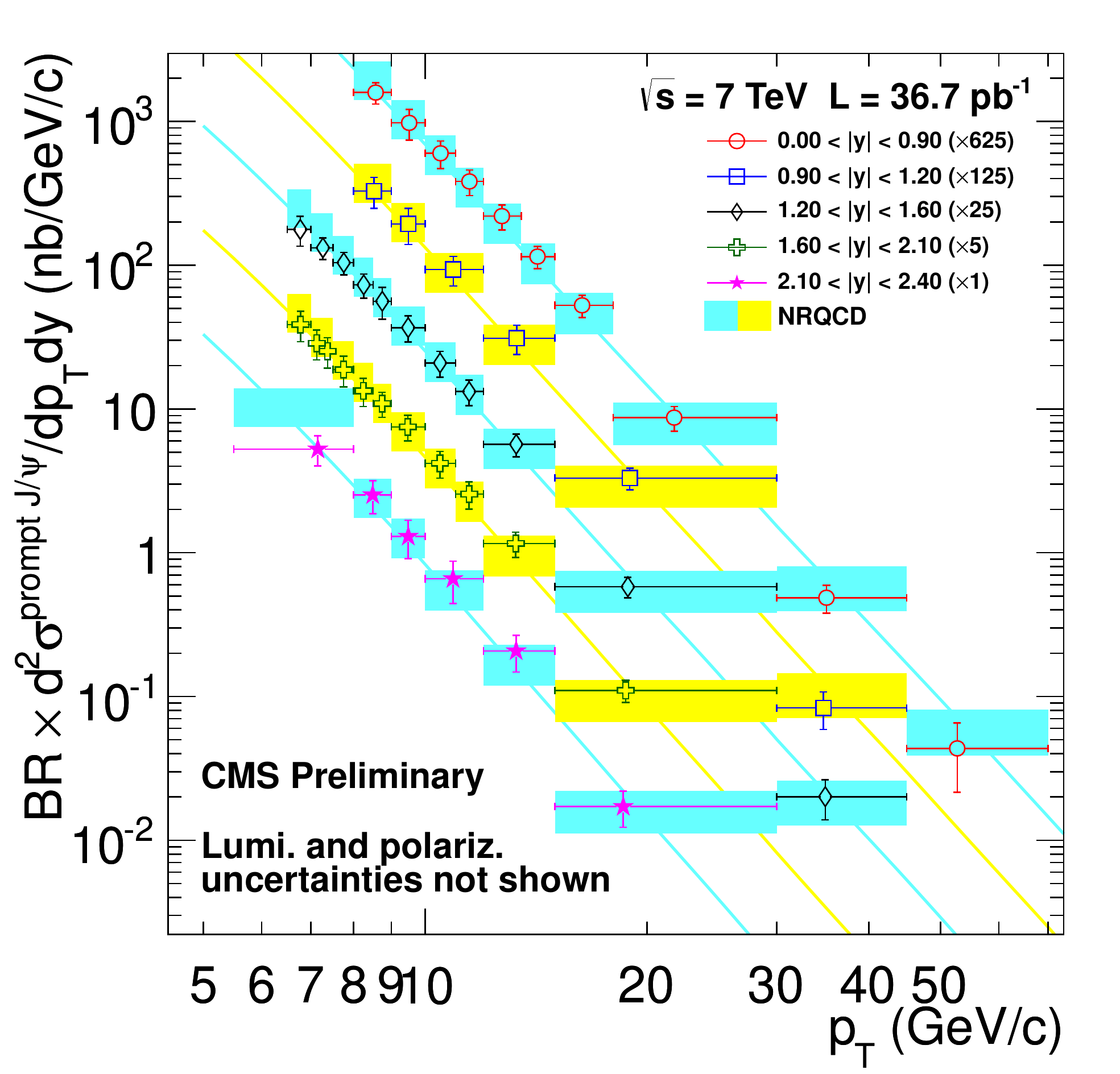}
\caption{Differential $J/\psi$ cross section measurement at CMS~\cite{Khachatryan:2010yr}.
\label{Fig:CMS_BPH10014_XS_allp_sg_jpsi_teor}}
\end{figure}

The prompt $J/\psi$ production has been measured by all four 
experiments~\cite{Khachatryan:2010yr,Aad:2011sp,Aaij:2011jh,Aamodt:2011gj}
with 2010 data in bins of rapidity and transverse momentum 
(See Fig.~\ref{Fig:CMS_BPH10014_XS_allp_sg_jpsi_teor}) and compared with theoretical models.
No large discrepancies are seen with the present level of uncertainties.
The unknown polarisation is the main uncertainty in all measurements,
and more data is needed to be able to resolve it.
LHCb also reports a cross-section for double $J/\psi$ production,
which is very sensitive to the production mechanism~\cite{Aaij:2011yc}.

\subsection{Open Charm}
The open charm cross-section has been measured using 
$D^0$, $D^+$, $D_s$ and $D^{\ast+}$ modes by Atlas and 
LHCb~\cite{LHCb-CONF-2010-013,ATLAS-CONF-2011-017}
using the first few $\rm nb^{-1}$ delivered by the LHC in 2010. The low luminosity
during this period allowed
to profit from a very open trigger which helps keeping systematic errors low.
The unexpected high total $c\bar c$ cross section of $6.10 \pm 0.93\rm\: mb$, 
extrapolated from these measurements is very encouraging for charm physics 
in $7\:$TeV collisions.
This is about 10\% of the total inelastic cross-section.

\subsection{Beauty}
The inclusive production of beauty and charm hadrons in pp collisions has been
measured by LHCb. In particular using semi-leptonic 
decays $b\to D^0(K\pi)\mu\bar{\nu}X$~\cite{Aaij:2011ju}
the cross section $\sigma(pp \to b\bar bX) = 284 \pm 20 \pm 49 \mu b$
is obtained~\cite{Bediaga:2010gn}, extrapolating to the
full phase space. All species of beauty hadrons can be produced in pp 
collisions, including $b$ baryons~\cite{LHCb-CONF-2011-001,LHCb-CONF-2011-036} 
and $B_c^+$~\cite{LHCb-CONF-2011-017}. 

The knowledge of the relative fractions of the various
$b$ hadron species is of crucial importance for all measurements of
branching rations, most prominently for $B_s\to\mu\mu$. LHCb have
measured the $B_s$ to $B_d$ production ratio using semileptonic
$B$ meson decays~\cite{LHCb-CONF-2011-028} and $SU(3)$ partner decays 
$B_d\to DK$ and $B_s\to D_s\pi$~\cite{Aaij:2011hi}. Both measurements are consistent
and get an average of $f_s/f_d=0.267\aerr{0.021}{0.020}$ \cite{LHCb-CONF-2011-034-001}.

\begin{figure}
\includegraphics[width=\columnwidth]{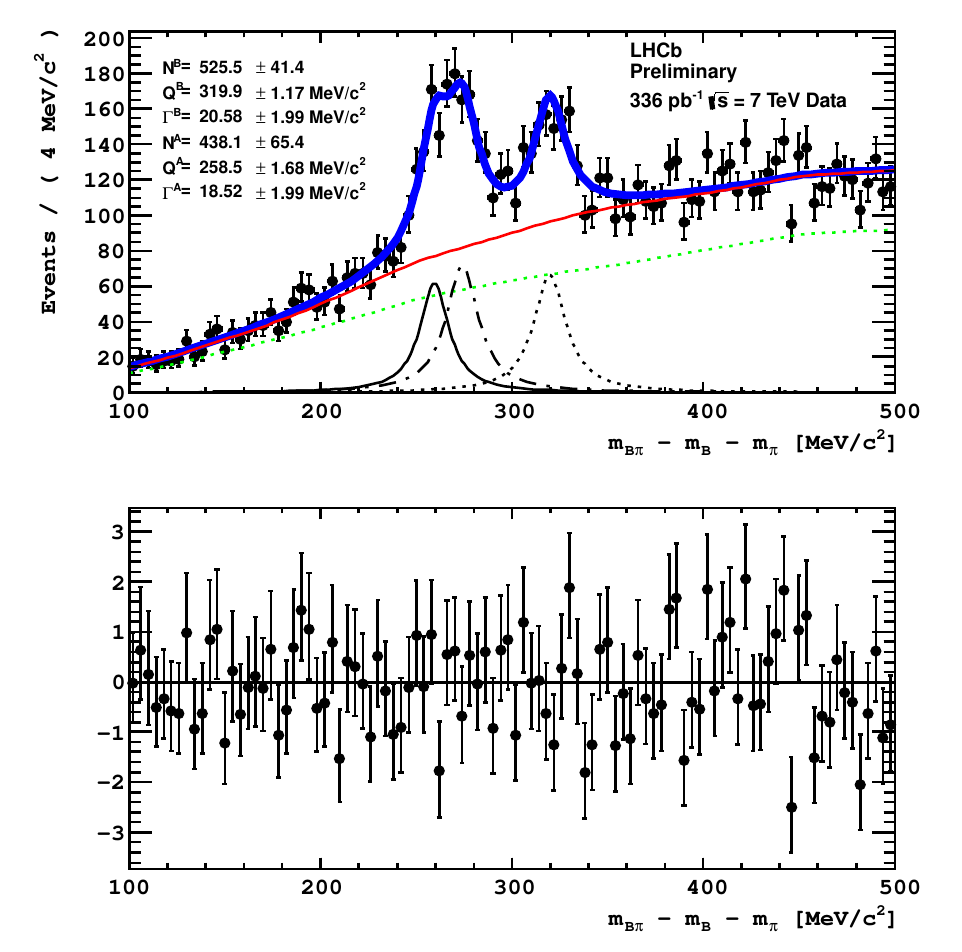}
\caption{Invariant mass relative to threshold of $B^0\pi$ system 
  ($m_{B^0\pi}-m_B-m_\pi$, top) and normalised fit residuals (bottom)~\cite{LHCb-CONF-2011-053}.
\label{Fig:ExcitedFig4a}}
\end{figure}
LHCb also studies orbitally excited $B$ mesons, notably observing
for the first time states decaying to $B^0\pi^+$ 
(Fig~\ref{Fig:ExcitedFig4a})~\cite{LHCb-CONF-2011-053}.
\section{Flavour Physics}
The LHCb experiment has been optimised for flavour physics
at the LHC and therefore all the results presented in this Section 
have been obtained by this experiment, with the notable exception
of the $B_s\to\mu\mu$ result from CMS.
\subsection{Charm Mixing}
The most interesting topic of charm physics is the characterisation
of neutral $D$ meson mixing and the hunt for $CP$ violation in 
$D$ meson decays. As for any other neutral long-lived meson (e.g. $K^0$, $B^0$, $B_s^0$),
the neutral $D$ system can be described in terms of two flavour eigenstates
$D^0$, $\overline{D}^0$ or two mass eigenstates:
$$
D_{1,2} = p\left|D^0\right> \pm q\left|\overline{D}^0\right>
$$
of masses $m_{1,2}$ and decay widths $\Gamma_{1,2}$. This allows to define the 
quantities
$$
  x = \frac{m_2-m_1}{2\Gamma}\qquad\text{and}\qquad y = \frac{\Gamma_2-\Gamma_1}{2\Gamma}.
$$
The HFAG averages for these quantities~\cite{Asner:2010qj} differ from the no mixing
hypothesis $x=0,y=0$ by $10.2\sigma$ but no single 
measurement excludes this hypothesis at $5\sigma$. 

Using two-body $D$ decays selected in $26\:\rm pb^{-1}$ of 2010 data the LHCb experiment measures
a linear combination of these quantities as~\cite{LHCb-CONF-2011-054}
\begin{eqnarray*}
  y_{CP}  & = & \frac{\hat\Gamma(D^0\to K^-K^+)}{\hat\Gamma(D^0\to K^-\pi^+)}-1\\
          & = & y\cos\phi -x\sin\phi\left(\frac{A_m}{2}+A_{\text{prod}}\right) \\
          & = & \left(-0.55\pm0.63\pm0.41\right)\%,
\end{eqnarray*}
where $1+A_m=|q/p|$ and the production asymmetry $A_{\text{prod}}$ is measured to be very small. 
In the limit of vanishing $CP$ violation $y_{CP}=y$. 
Using the same data sample, LHCb also measure the lifetime difference~\cite{LHCb-CONF-2011-046}
\begin{eqnarray*}
A_\Gamma & = & \frac{\tau(\overline{D}^0\to K^+{}K^-)-\tau(D^0\to K^+{}K^-)}{\tau(\overline{D}^0\to K^+{}K^-)+\tau(D^0\to K^+{}K^-)} \\
         & = & \left(-0.59\pm0.59\pm0.21\right).
\end{eqnarray*}
\begin{figure}
\includegraphics[width=0.48\columnwidth]{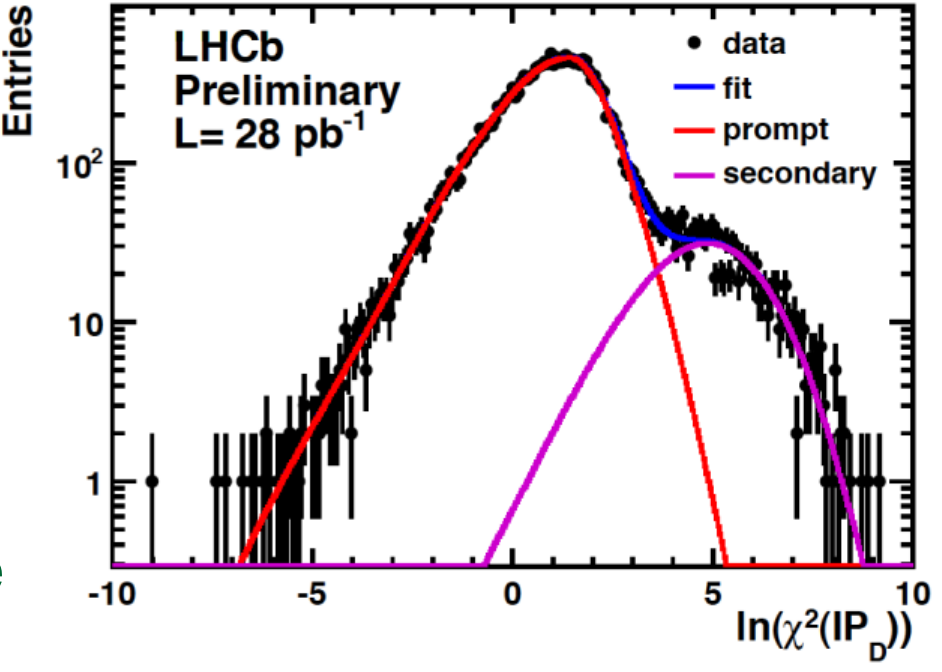}
\includegraphics[width=0.48\columnwidth]{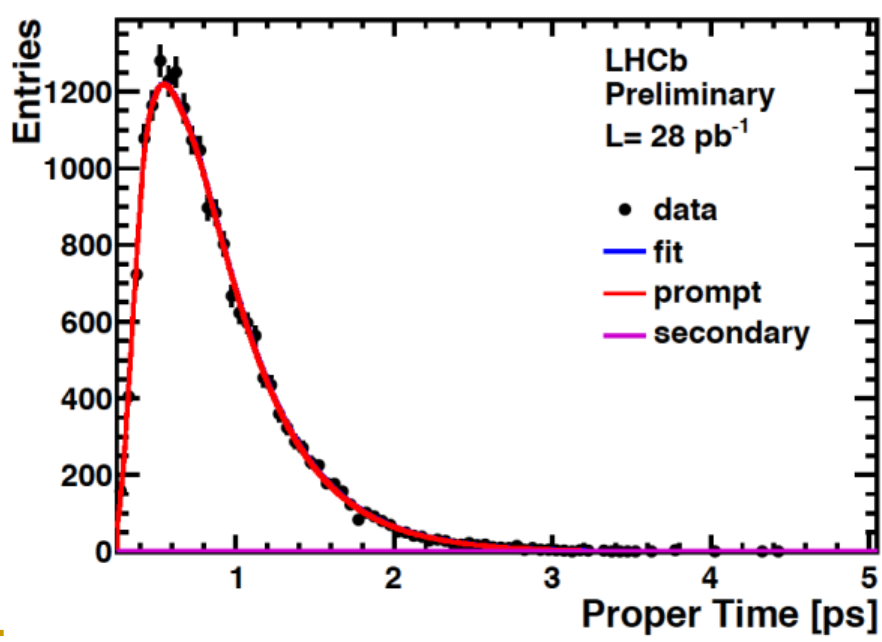}
\caption{$D^0$ impact parameter (left) and proper time~\cite{LHCb-CONF-2011-046}\label{Fig:Charm}}
\end{figure}
The key to the lifetime measurement is a good separation of prompt and secondary 
charm (from $b$ decays), illustrated in Fig~\ref{Fig:Charm}.

\subsection{$CP$ Violation in Charm}
\begin{figure}
\includegraphics[width=0.48\columnwidth]{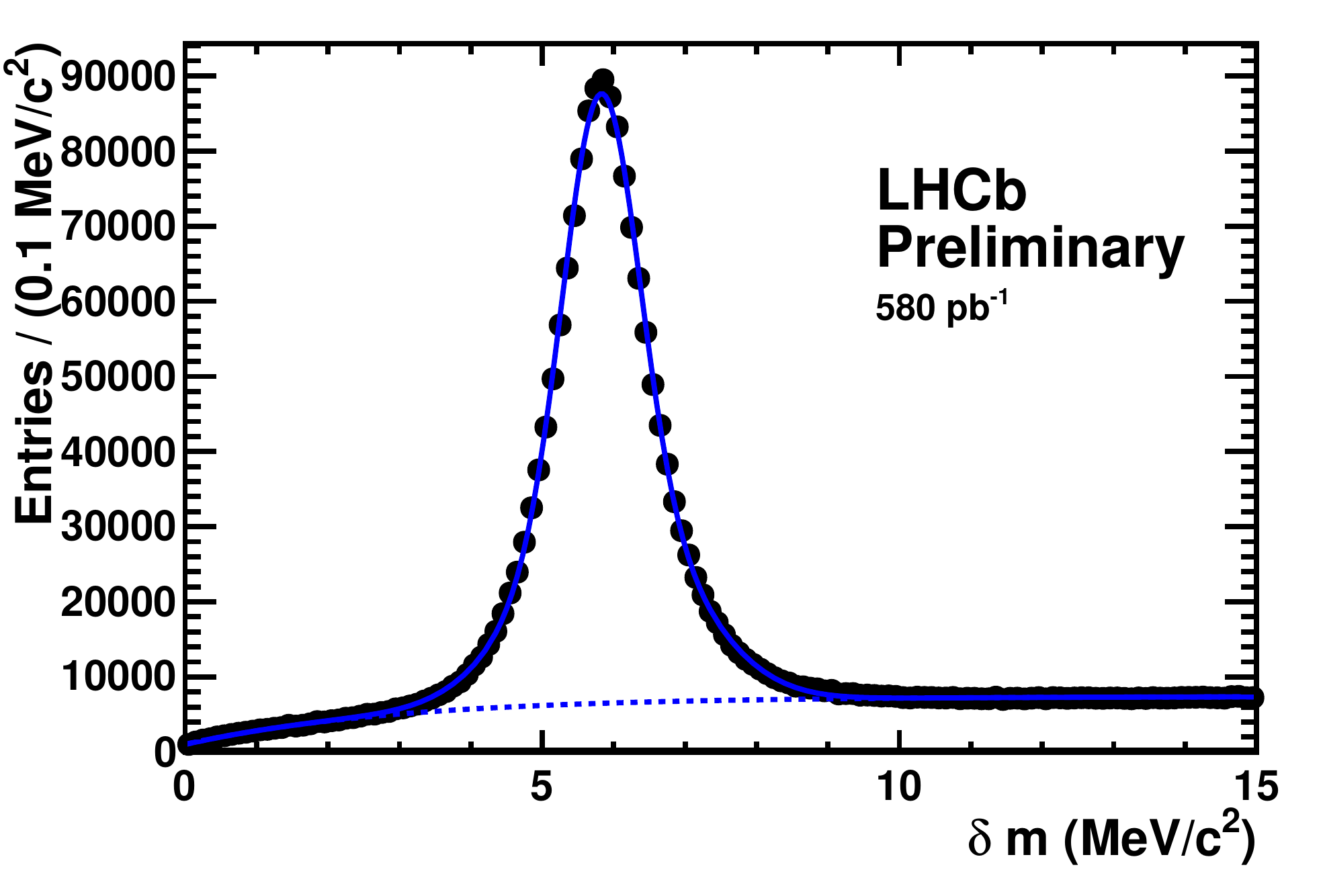}
\includegraphics[width=0.48\columnwidth]{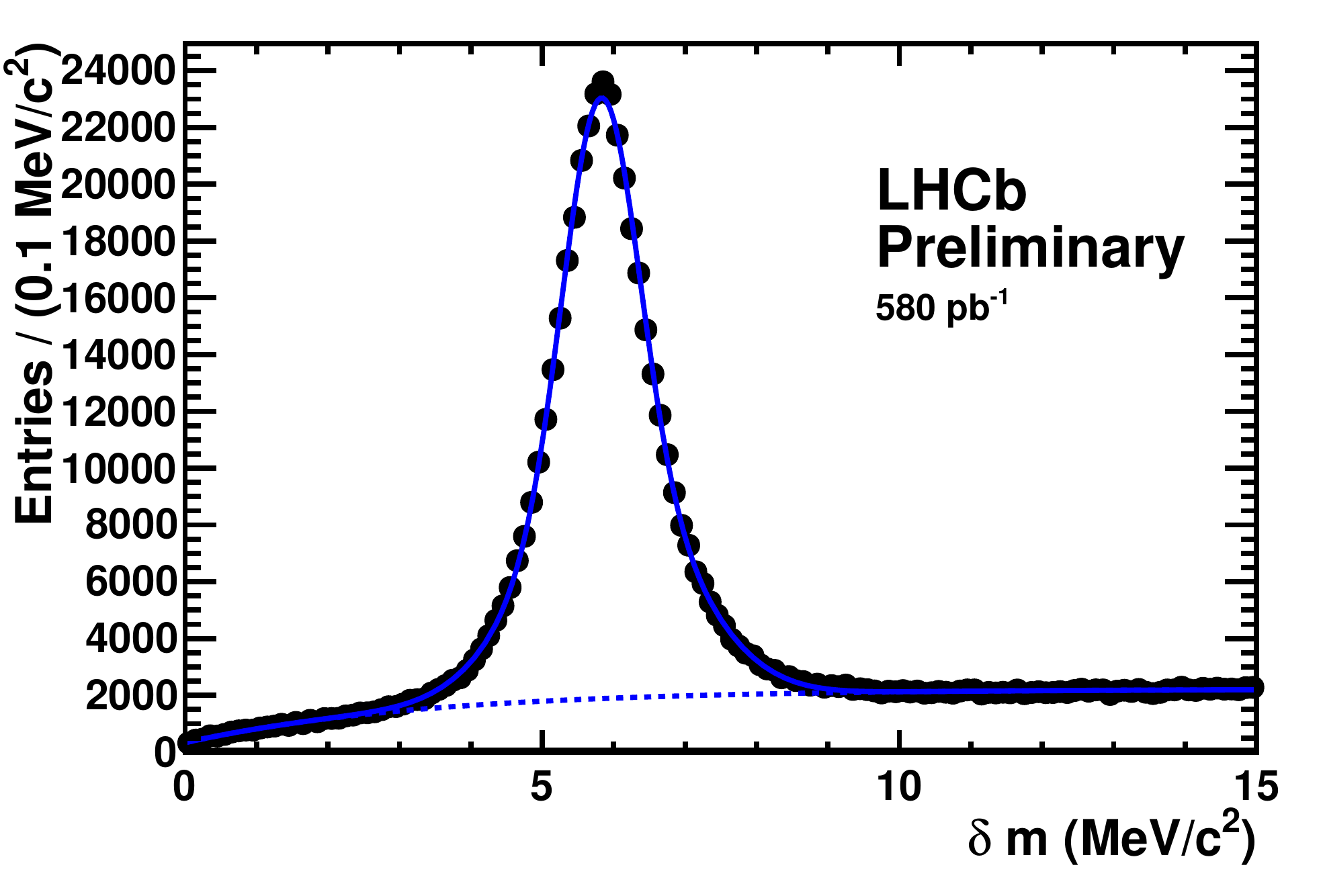}
\caption{Fits to the mass difference spectrum in tagged $D\to K^+K^-$ (left) 
and $D\to\pi^+\pi^-$ (right) decays~\cite{LHCb-CONF-2011-061}\label{Fig:DeltaAcp}}
\end{figure}

$CP$ violation is expected to be vanishingly small in the charm sector in the Standard Model.
A non-zero $CP$ asymmetry above a few per-mille in $D^0\to h^+ h^-$ ($h=\pi,K$)
decays would be strong sign of new physics. Experimentally the flavour of the $D$ meson is tagged
using the decay $D^{\ast}\to D^0\pi^+$. 
The raw $CP$ asymmetry of
tagged $D^0\to f$ and $\overline{D}^0\to f$ can be factorised as
$$
 A_{\text{RAW}}(f) = A_{CP}(f) +  A_{\text{D}}(f) +  A_{\text{D}}(\pi_s) +  A_{\text{P}}(D^{\ast})
$$
where $A_D$ are detector asymmetries related to the final state $f$ and the 
bachelor pion $\pi_s$ and $A_{\text{P}}(D^{\ast})$ is the production asymmetry in 
$pp$ collisions. The detection asymmetry for $f$ vanishes when one uses decays
to $CP$ eigenstates, e.g. $\pi^+\pi^-$ or $K^+K^-$. 
All other asymmetries can be cancelled at first order by measuring the difference 
of the two $CP$ asymmetries in these two channels. 
While writing these proceedings the following interesting result has become available.
Using $580\:\rm pb^{-1}$ of 2011 data the LHCb collaboration 
gets very clean samples of $1.44\cdot10^6$ tagged $D\to K^+K^-$ and
$0.38\cdot10^6$ $D\to\pi^+\pi^-$ 
(Fig~\ref{Fig:DeltaAcp})~\cite{LHCb-CONF-2011-061}.
Due to the different lifetime acceptance of
the two channels a small contribution from mixing induced $CP$ violation
does not cancel out in the measurement but its magnitude can be extracted from data:
\begin{eqnarray*}%
\Delta A_{CP} & \equiv & A_{CP}^{\rm raw}(K^+K^-)-A_{CP}^{\rm raw}(\pi^+\pi^-) \\ 
               &=& A_{CP}^{\rm dir}(K^+K^-)-A_{CP}^{\rm dir}(\pi^+\pi^-) + 0.098\:A_{CP}^{\rm ind}\\ 
               &=&\left(\-0.82\pm0.21\pm0.11\right)\%.
\end{eqnarray*}%
The measured difference of $CP$ asymmetries
the first ($3.5\sigma$) evidence of 
$CP$ violation in the charm sector. 

\subsection{Rare $b$ Decays}
The SM prediction for the Branching Ratios (BR) of the decays $B_q\to\mu^+\mu^-$
have been computed to be ${\rm BR}(B_s\to\mu^+\mu^-)= (3.2 \pm0.2)\cdot10^{-9}$
and ${\rm BR}(B_d\to\mu^+\mu^-)= (0.10\pm0.01)\cdot10^{-9}$~\cite{Buras:2010wr}. 
However, many extensions of the SM predict large
enhancements to these BR. The first search for this at the LHC decay was 
reported by the LHCb collaboration with 2010 data~\cite{Aaij:2011rja}.
Recently LHCb and CMS collaborations have presented new searches
based on $0.3$ and $1.1\:\rm fb^{-1}$ samples collected in 2011,
respectively~\cite{LHCb-CONF-2011-037,Chatrchyan:1371756}.

\begin{figure}
\includegraphics[width=\columnwidth]{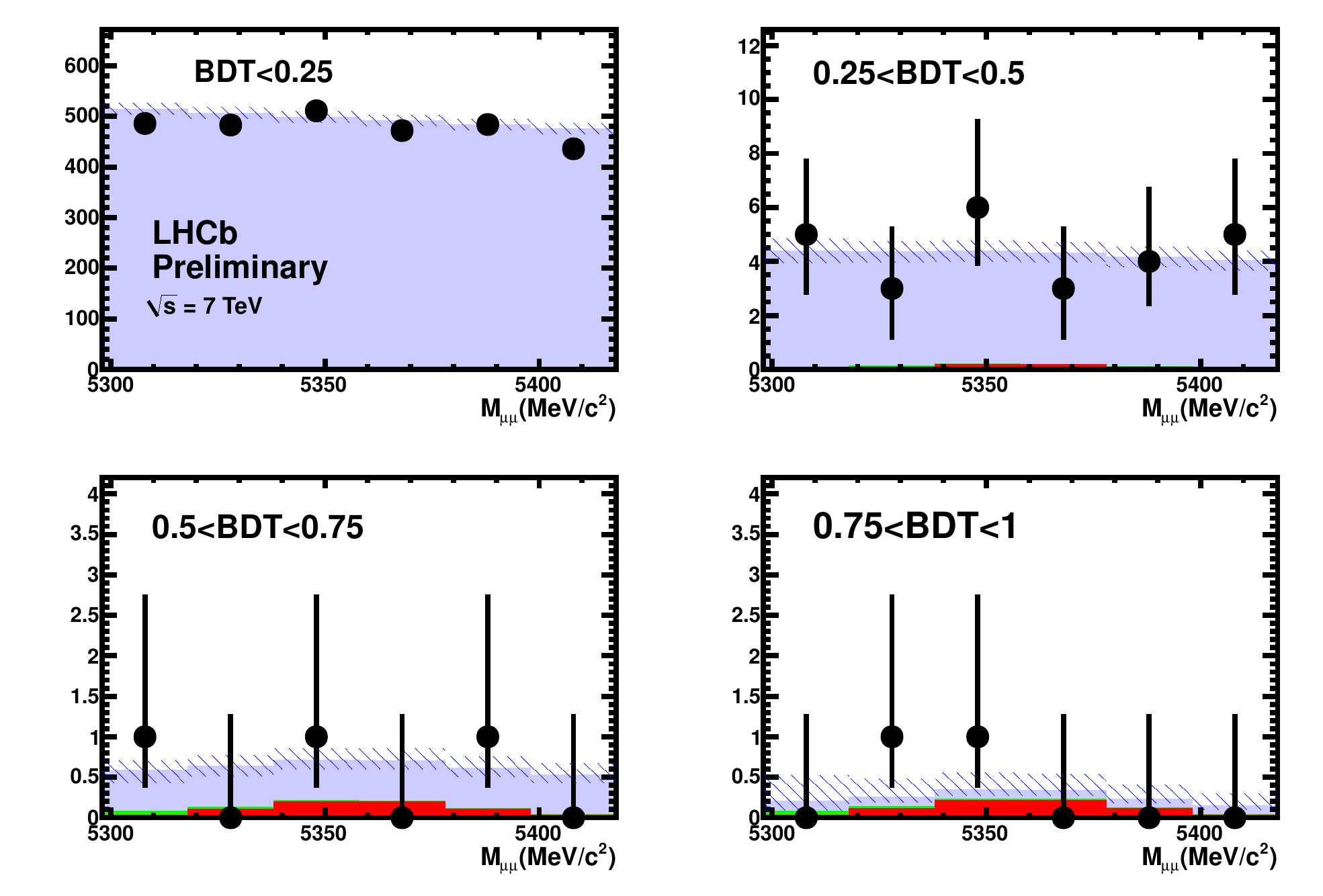}
\includegraphics[width=\columnwidth]{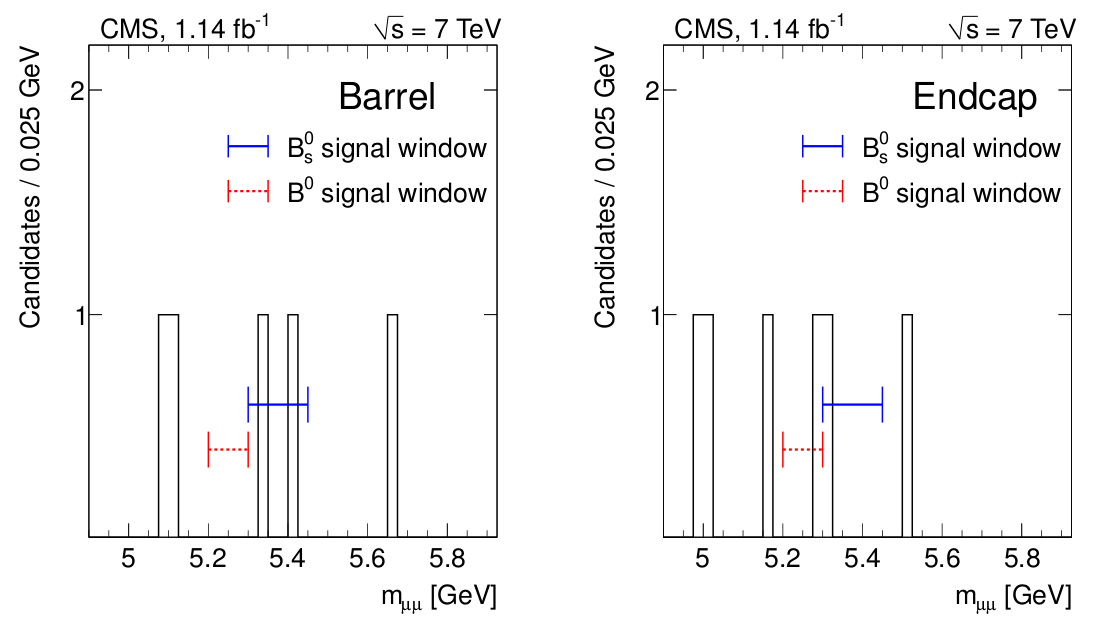}
\caption{$B_s\to\mu\mu$ search windows at LHCb~\cite{LHCb-CONF-2011-037} (top) 
  and CMS~\cite{Chatrchyan:1371756} (bottom).\label{Fig:Bsmm}}
\end{figure}
While CMS uses a cut-based approach, LHCb uses a boosted decision tree calibrated
on $B\to hh$ ($h=\pi,K$) decays which have the same topology as the signal. 
The estimated yield is then normalised using $B_d\to K\pi$ (LHCb only), 
$B^+\to J/\psi K$ and $B_s\to J/\psi\phi$. 
No excess of signal is observed at neither of the two experiments (Fig.~\ref{Fig:Bsmm}) and
upper limits are set by LHCb as ${\rm BR}(B_{s}\to\mu^+\mu^-)<1.6\cdot10^{-8}$ (95\% C.L.) and 
${\rm BR}(B_{d}\to\mu^+\mu^-)<5.1\cdot10^{-9}$, and CMS as ${\rm BR}(B_{s}\to\mu^+\mu^-)<1.8\cdot10^{-8}$.
The combined LHC result is ${\rm BR}(B_{s}\to\mu^+\mu^-)<1.1\cdot10^{-8}$~\cite{CMS-PAS-BPH-11-019} 
thus not confirming
the excess reported by CDF~\cite{Aaltonen:2011fi}.
The limits set by the LHC strongly constrain the allowed SUSY parameter 
space, especially at large $\tan\beta$~\cite{Akeroyd:2011kd}.

The rare decay $B_d\to\mu\mu K^\ast$ is a $b\to s$ flavour changing neutral current decay
which is in the SM mediated by electroweak box and penguin diagrams. It can be a
highly sensitive probe for new right handed currents and new scalar and pseudoscalar
couplings. These New Physics contributions can be probed by its contribution to the angular distributions
of the $B^0$ daughter particles. The most prominent observable is the forward-backward
asymmetry of the muon system ($A_{\rm FB}$). $A_{\rm FB}$ varies with the invariant mass-squared
of the dimuon pair ($q^2$) and in the SM changes sign at a well defined point, where
the leading hadronic uncertainties cancel. In many NP models the shape of $A_{\rm FB}$ as
a function of $q^2$ can be dramatically altered.
\begin{figure}
\includegraphics[width=\columnwidth]{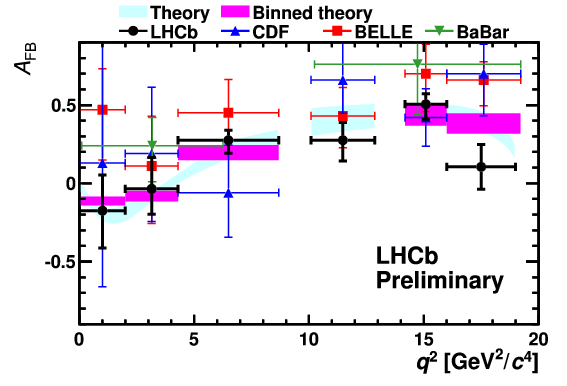}
\includegraphics[width=\columnwidth]{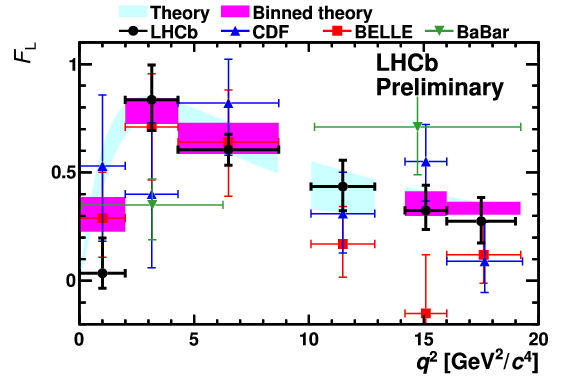}
\caption{$B_d\to\mu\mu K^\ast$ forward-backward asymmetry $A_{\rm FB}$ (top)
  and $K^\ast$ polarisation fraction
  $F_L$ in bins of dimuon mass $q^2$~\cite{LHCb-CONF-2011-038}.\label{Fig:BmmKs}}
\end{figure}
The latest LHCb analysis~\cite{LHCb-CONF-2011-038} uses $309\rm\:pb^{-1}$ of data collected 
during 2011 to measure $A_{\rm FB}$, the fraction of longitudinal polarisation of the $K^\ast$,
$F_L$, and the differential branching fraction, $dB/dq^2$, as a function of the dimuon
invariant mass squared, $q^2$. There is good agreement between recent Standard Model
predictions and the LHCb measurement of $A_{\rm FB}$, $F_L$ and $dB=dq^2$ in the six $q^2$ bins
(Fig.~\ref{Fig:BmmKs}).
In a $1 < q^2 < 6 \rm\: GeV^2$ bin, LHCb measures $A_{\rm FB} = 0.10\pm0.14\pm0.05$, to be compared
with theoretical predictions of $A_{\rm FB} = 0.04 \pm 0.03$. The experimental uncertainties
are presently statistically dominated, and will improve with a larger data set. Such
a data set would also enable LHCb to explore a wide range of new observables.

Using a very similar selection, LHCb also searched for Majorana Neutrinos~\cite{Aaij:2011ex} giving raise
to $B^+\to K^-\mu^+\mu^+$ and $B^+\to\pi^-\mu^+\mu^+$ decays. No excess was found 
and 95\% C.L. limits have been set at $5.4\cdot10^{-8}$ and $5.8\cdot10^{-8}$, respectively.

\begin{figure}
\includegraphics[width=0.48\columnwidth]{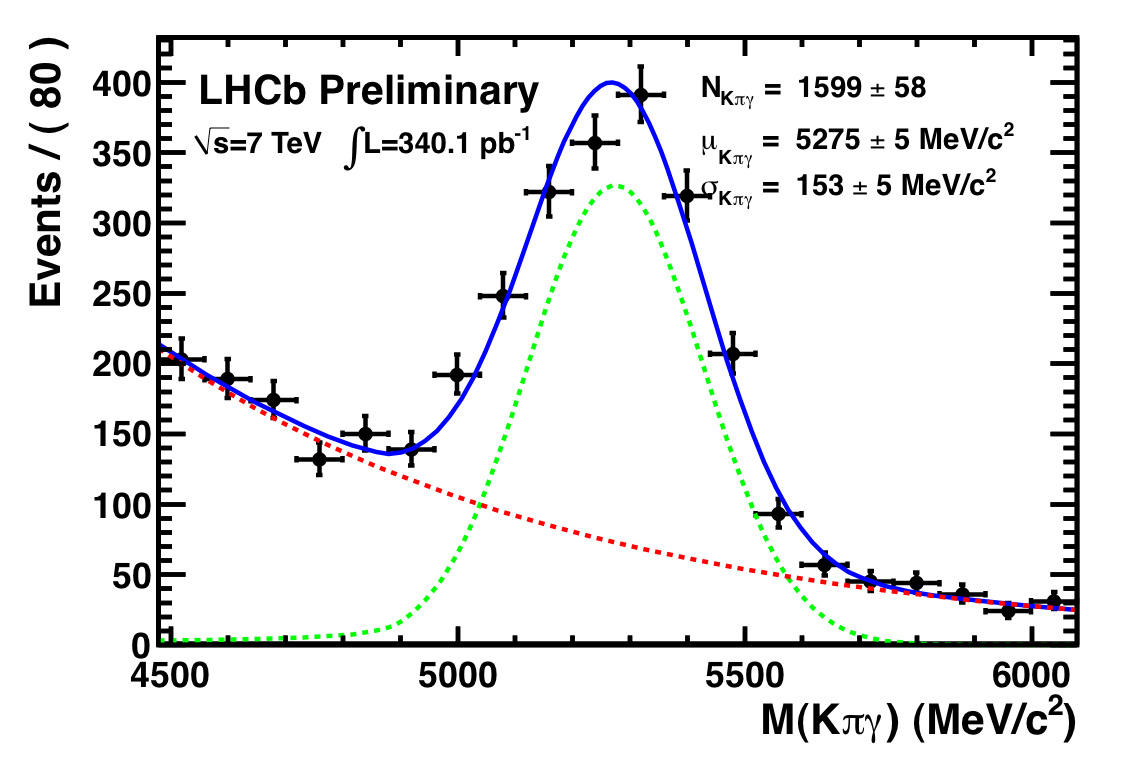}
\includegraphics[width=0.48\columnwidth]{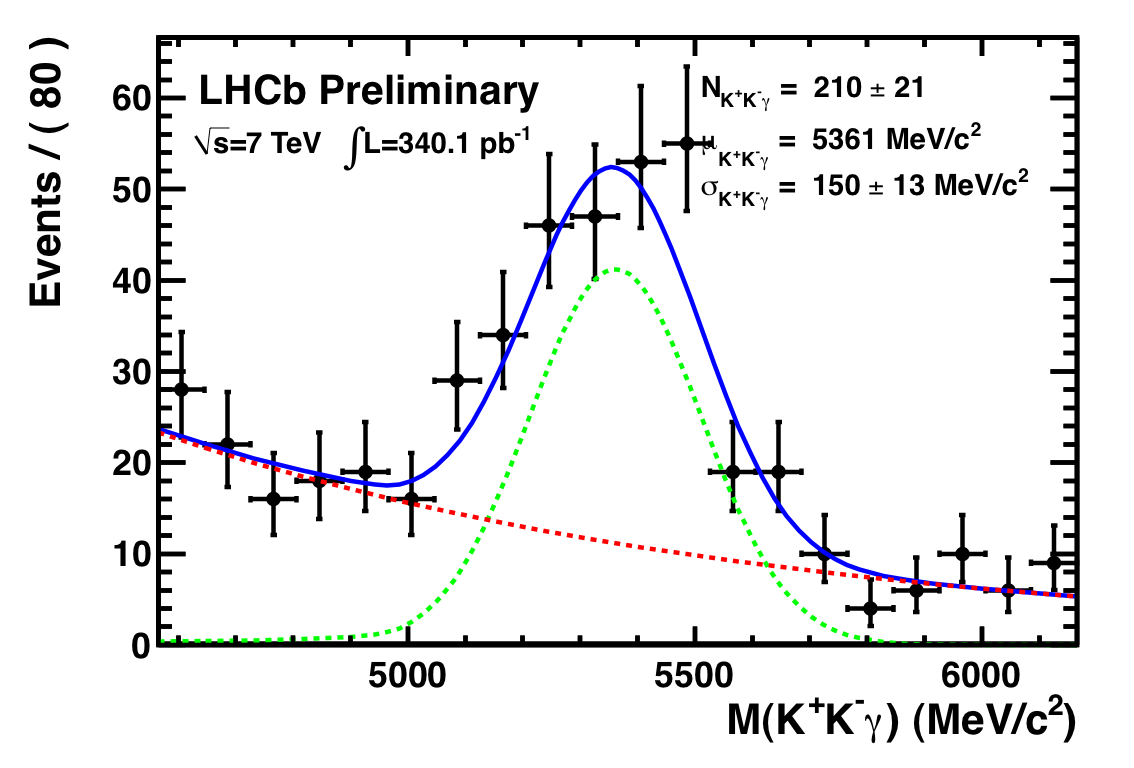}
\caption{$B\to K^\ast\gamma$ and $B_s\to \phi\gamma$ mass peaks~\cite{LHCb-CONF-2011-055}.
  \label{Fig:BsGamma}}
\end{figure}
Other $b\to s$ transitions of interest are radiative decays $B\to K^\ast\gamma$ and
 $B_s\to \phi\gamma$. LHCb reports the first measurement of the ratio of
branching fractions of these two decays as $1.52\pm0.15\pm0.10\pm0.12$ where 
the last error comes from ($f_d/f_s$)~\cite{LHCb-CONF-2011-055}.   
The mass resolution (Fig~\ref{Fig:BsGamma}) is dominated by the photon energy resolution.
LHCb already has the largest sample of $B_s\to \phi\gamma$, which will become to measure
or constrain non-standard right-handed currents.

\subsection{CP violation in $B$ decays}
Decays of neutral $B$ mesons provide a unique laboratory to study CP-violation originating
from a non-trivial complex phase in the CKM matrix. The relative phase
between the direct decay amplitude and the amplitude of decay via mixing gives rise
to time-dependent CP-violation, a difference in the proper decay time distribution of
$B$-meson and anti-$B$-meson decays. The decay $B_s\to J/\psi\phi$
is considered the golden modes for measuring this type of CP-violation,
In the Standard Model the CP-violating phase in this decay is predicted to be 
$\phi_{s}\simeq-2\beta_s$ where $\beta_s = \arg(-V_{ts}V_{tb}^\ast/V_{cs}V_{cb}^\ast)$.
The indirect determination via global fits to experimental data gives 
$2\beta_s = (0.0363\aerr{0.0016}{0.0015})\:\rm rad$. New Physics contributions could
significantly alter this phase. 

The channel $B_s\to J/\psi f_0(980)$ is also sensitive
to the same phase. It has been first observed by the LHCb collaboration~\cite{Aaij:2011fx}
using 2010 data and quickly confirmed by Belle~\cite{Li:2011pg} and CDF~\cite{Aaltonen:2011nk}.

\begin{figure}
\includegraphics[width=\columnwidth,clip=true,trim=0cm 5cm 0cm 3cm]{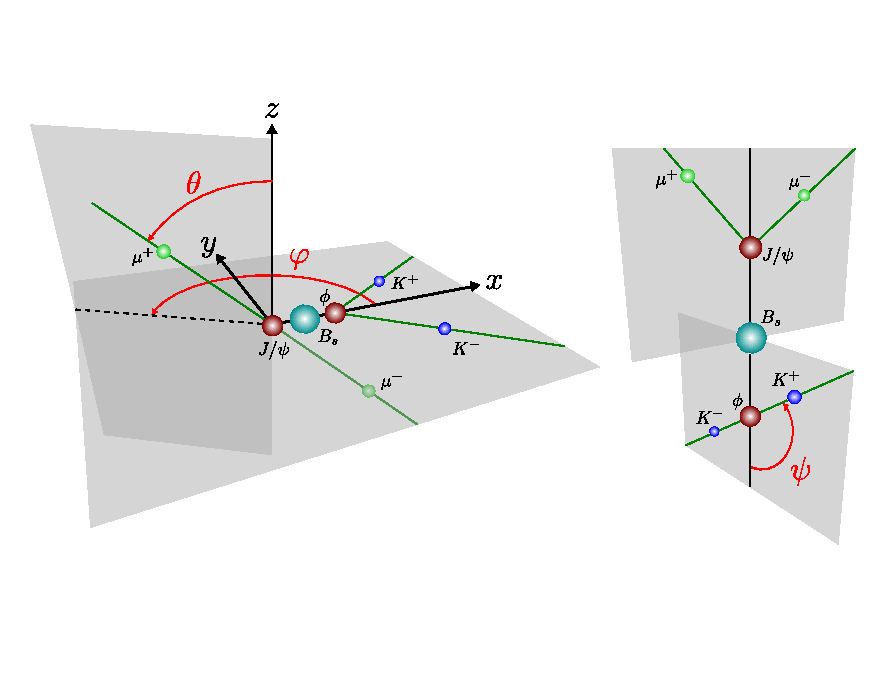}
\caption{Definitions of the decay angles in $B_s\to J/\psi\phi$ decay.
\label{Fig:BetasAnglesDef}}
\end{figure}

\begin{figure}
\includegraphics[width=0.48\columnwidth]{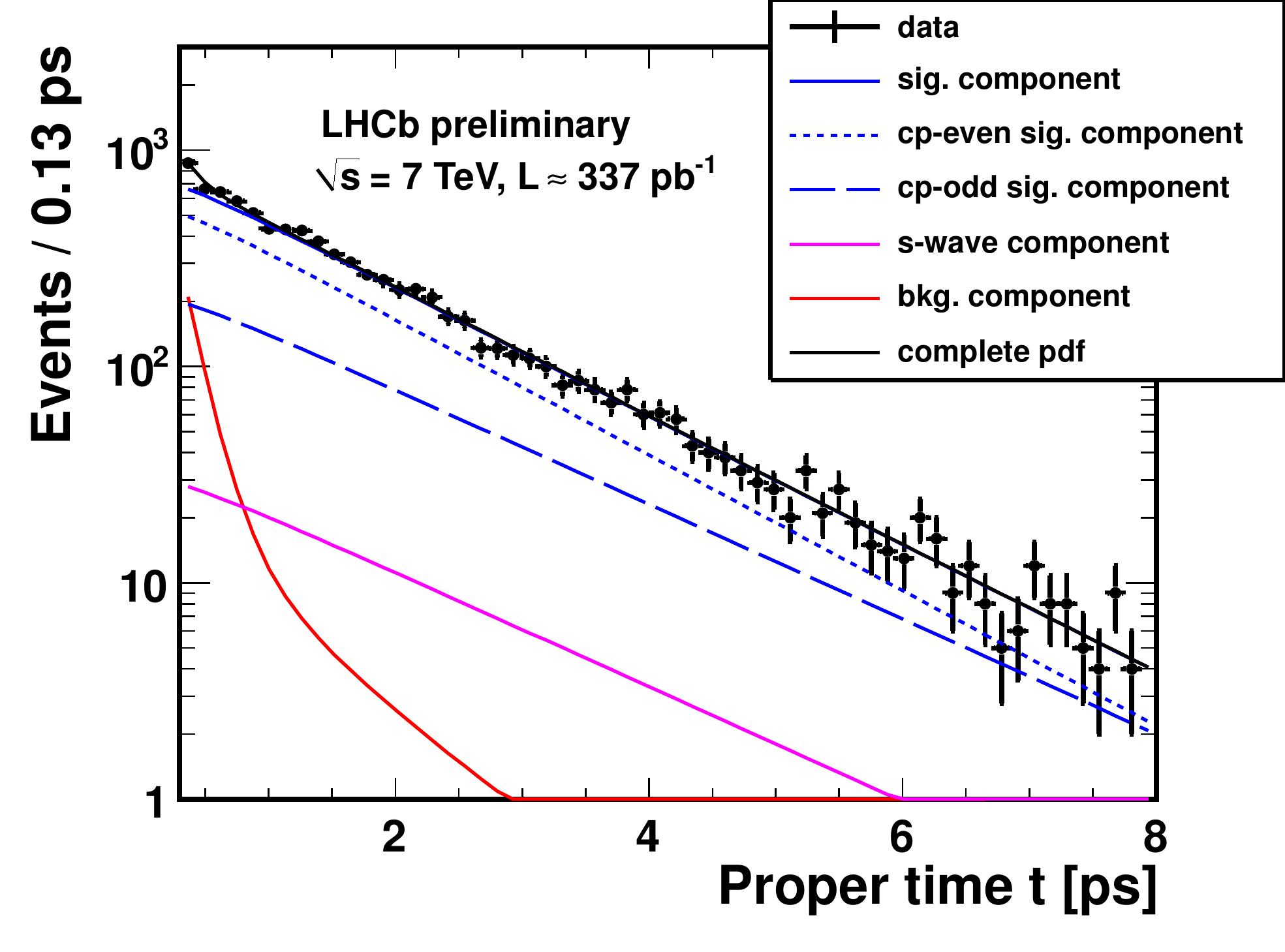}
\includegraphics[width=0.48\columnwidth]{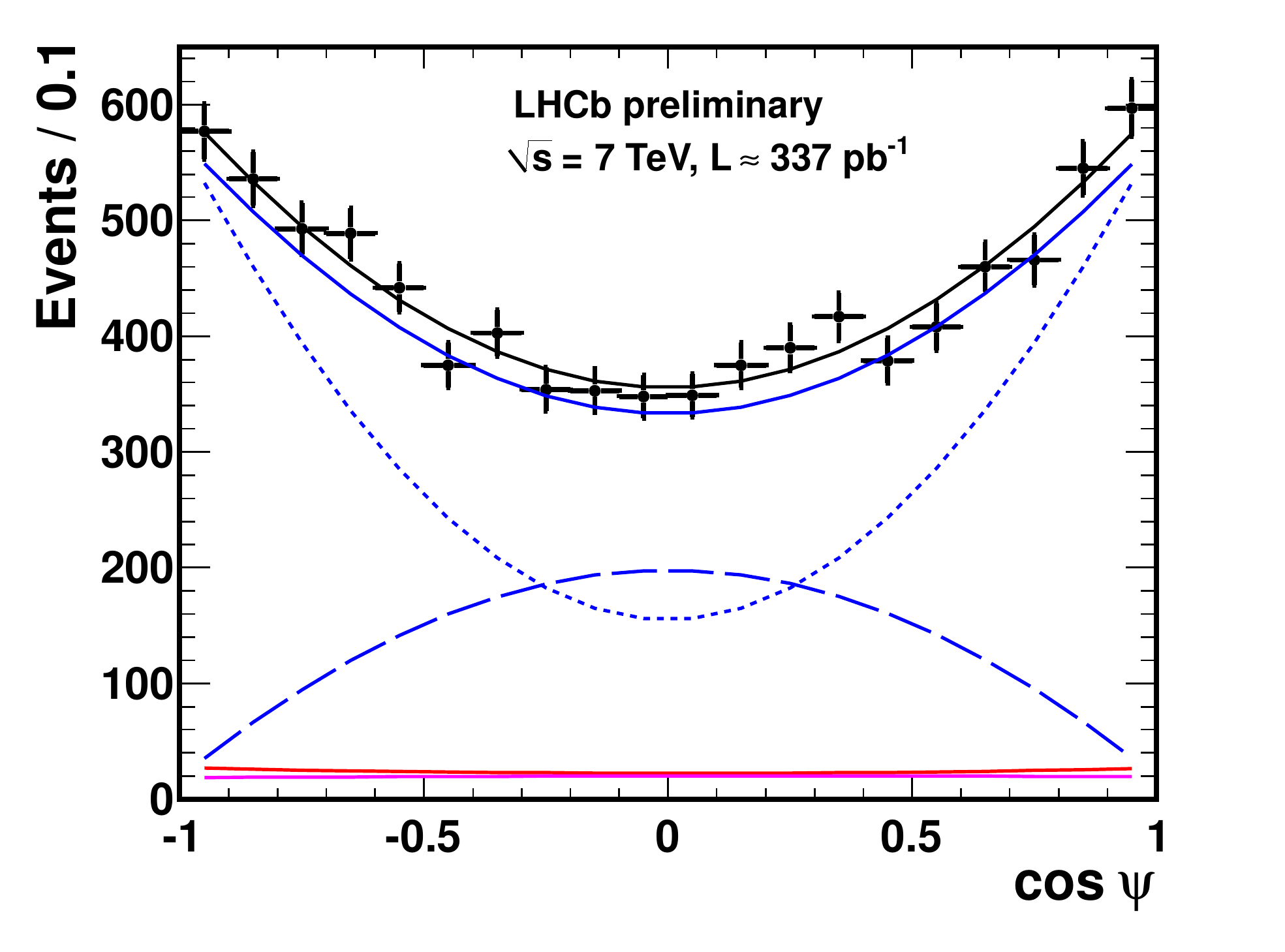}\\
\includegraphics[width=0.48\columnwidth]{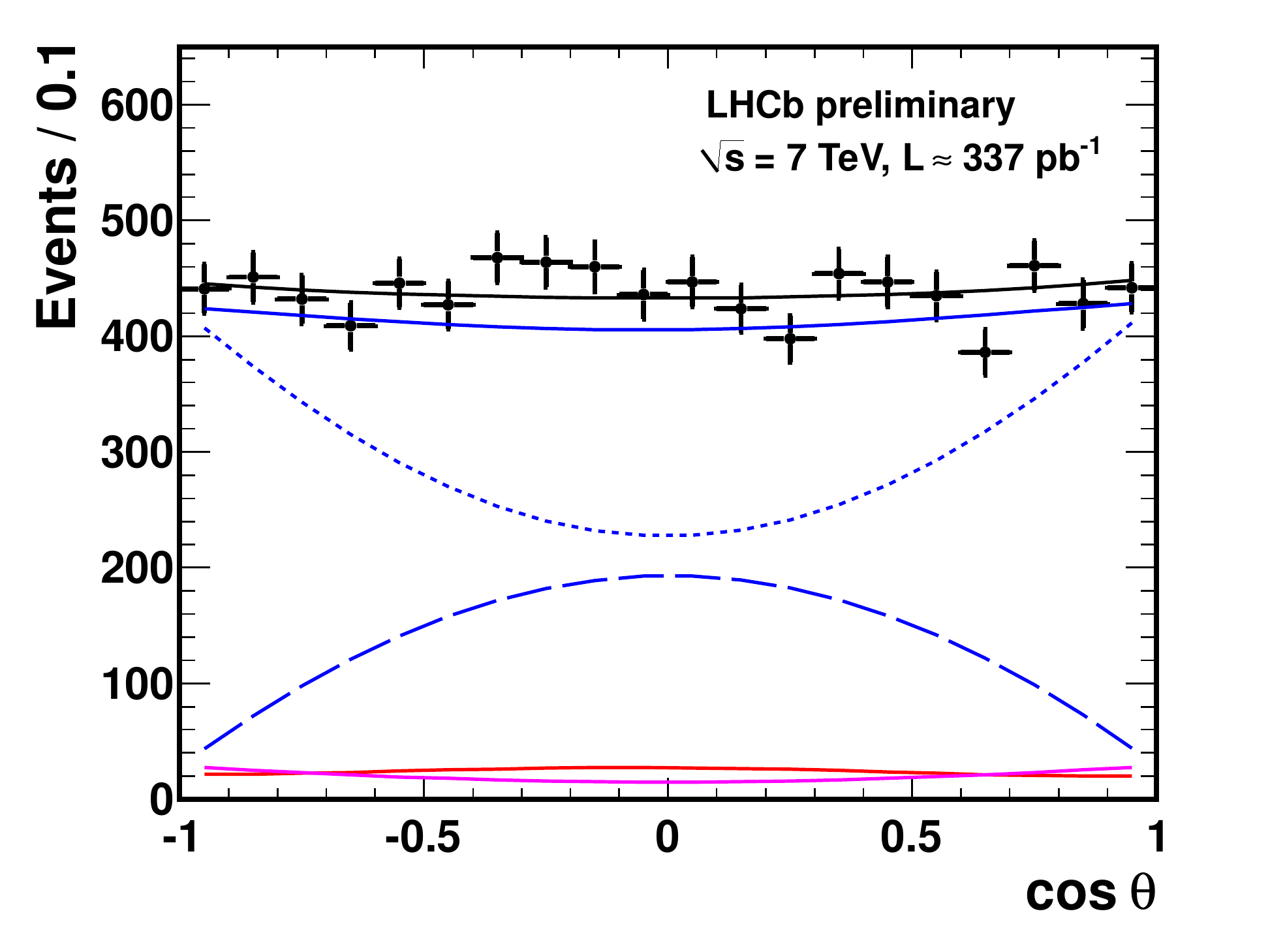}
\includegraphics[width=0.48\columnwidth]{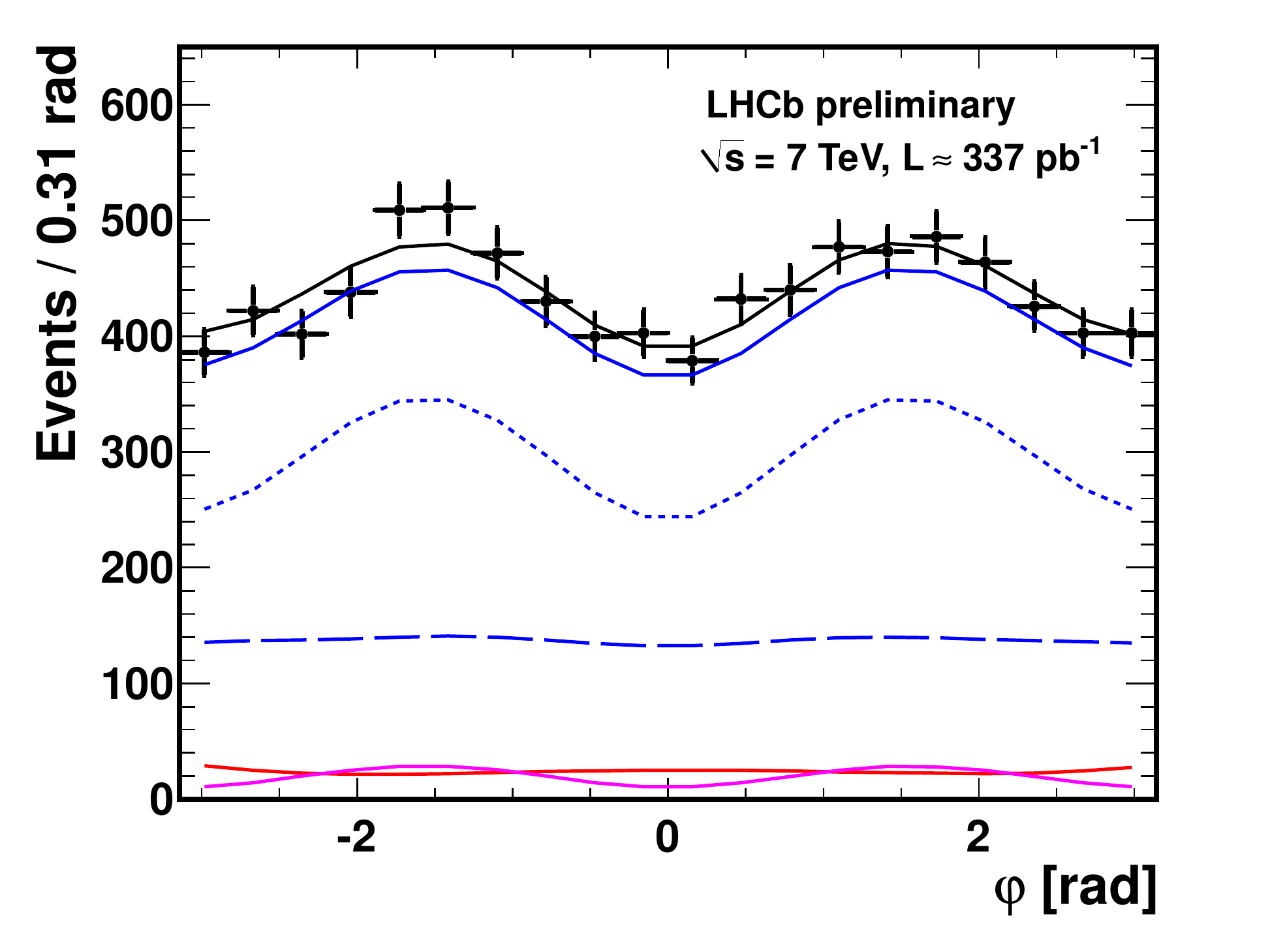}
\caption{Fits projections to proper time (top left),
 and the three angles in the $B_s\to J/\psi\phi$ decay~\cite{LHCb-CONF-2011-049}.
\label{Fig:BetasAngles}}
\end{figure}

LHCb report measurements of the phase $\phi_s$ for each of these channels using
$338\:\rm fb^{-1}$, and also performing a simultaneous fit to both channels. 
In both cases, flavour-tagged and untagged events are used, and the tagging efficiency
is calibrated to control channels. The trigger and selection bias, in particular
with respect to lifetime, is also extracted from the data itself. 
Due to the vector nature of the $\phi$ meson,
the $B_s\to J/\psi\phi$ needs an angular analysis to disentangle the CP-even and 
CP-odd final states (Figs.~\ref{Fig:BetasAnglesDef} and~\ref{Fig:BetasAngles}). 
This is not necessary in the $f_0(980)$ case.

\begin{figure}
\includegraphics[width=\columnwidth]{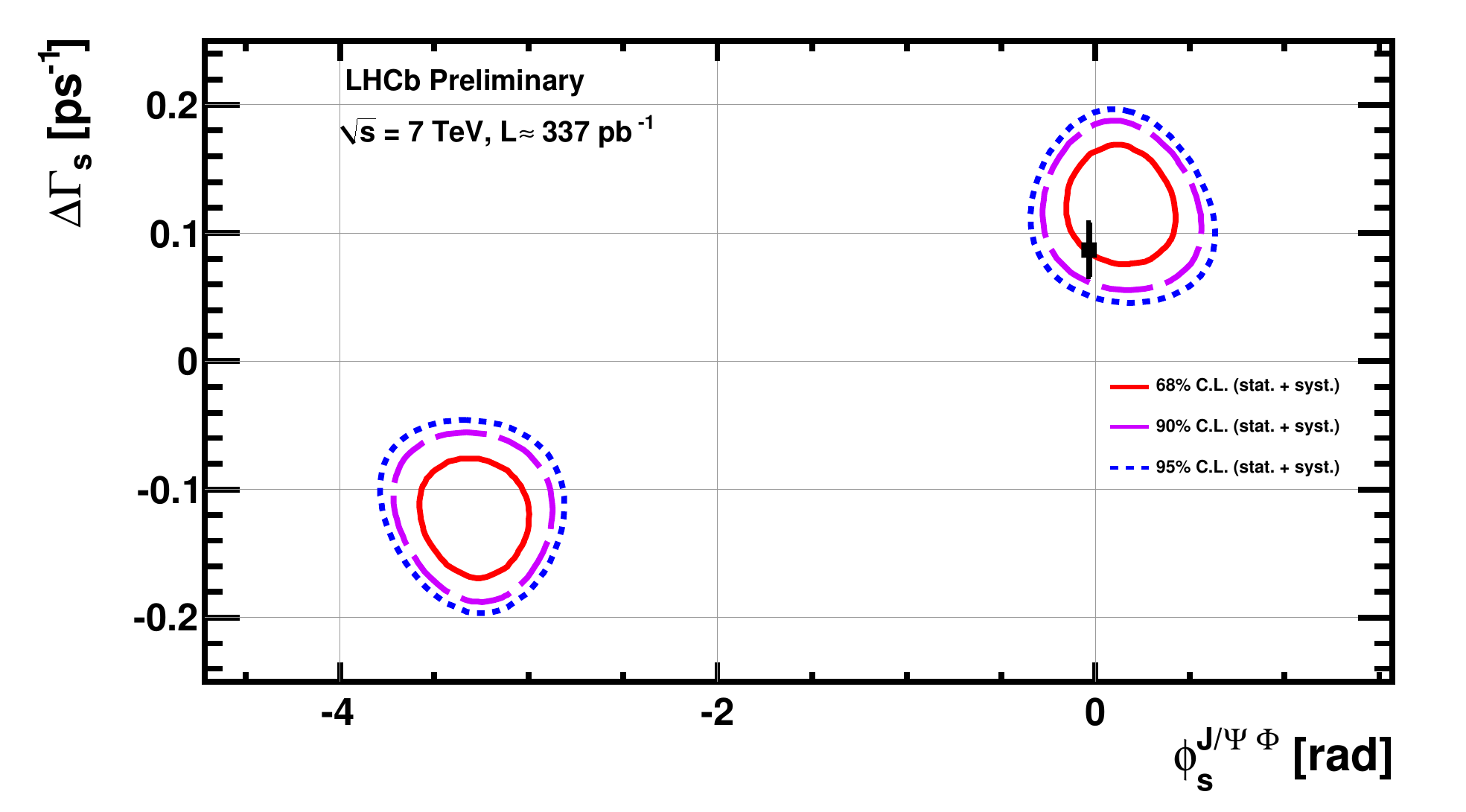}
\caption{Constraints of LHCb's measurement on the $\phi_s$--$\Delta\Gamma_s$ 
  plane from $B_s\to J/\psi\phi$~\cite{LHCb-CONF-2011-049}.
\label{Fig:Betas}}
\end{figure}

The results are~\cite{LHCb-CONF-2011-049,LHCb-CONF-2011-051,LHCb-CONF-2011-056}
\begin{eqnarray*}%
\phi_s^{J/\psi f_0} & = & -0.44 \pm 0.44 \pm 0.02~\mathrm{ rad} \\%
\phi_s^{J/\psi\phi} &=& +0.13\pm0.18\pm0.07~\mathrm{ rad}\\ %
\phi_s^{\rm Comb} &=& +0.03\pm0.16\pm0.07~\mathrm{ rad}  %
\end{eqnarray*}
which are consistent with the SM prediction. A sign ambiguity remains
under the sign reversal of ($\phi_s$) and ($\Delta\Gamma_s$). The allowed
regions are shown in Fig~\ref{Fig:Betas}.

With increasing precision on $CP$ violating phases in $b\to c\bar{c}s$ transitions,
assumed to be dominated by tree-level topologies,
it will become crucial in the future to understand contributions from
penguin topologies~\cite{Bs2JpsiKs,Bs2JpsiKs_conf}. These can be studied 
using Cabibbo-suppressed decays that are related by $U$-spin symmetry. One example
is the $B_s\to J/\psi K_S^0$ decay, which is the partner of the golden mode 
$B_d\to J/\psi K_S^0$. CDF~\cite{CDFBs} and LHCb~\cite{LHCb-CONF-2011-048} 
have recently reported on the branching ratio
of this channel and more precision will become available 
when more data is collected.

\section{Conclusions}
With its large $b$ and $c$ cross-sections, the LHC is the new flavour factory. 
Many flavour physics results, mostly from LHCb, are becoming available
yielding an unexpected and interesting pattern of measurements. 
The long awaited $B_s\to\mu\mu$ decay has not yet been observed, 
thus excluding large regions of the SUSY 
parameter space. Similarly the $CP$ violating
phase in $B_s$ decays is compatible with the SM expectation,
as well as the angular distributions in $B\to K^\ast\mu\mu$.
Yet all these measurements are still affected by statistical errors 
much larger than the theoretical errors, which leaves a lot of room
for observations of new physics.

The biggest surprise comes from the measurement of $\Delta A_{CP}$ 
which exhibits a $3.5\sigma$ evidence for new physics. This will all
have to be followed up very closely with increasing statistics.
As high-precision beauty and charm physics is sensitive to energy scales much beyond 
the LHC centre-of-mass energy it is likely that flavour physics
is paving the way for direct observations of new particles 
by the general purpose detectors at the LHC.



\bibliography{PIC}

\begin{thebibliography}{10}

\bibitem{Alves:2008zz}
LHCb Collaboration.
\newblock {The LHCb Detector at the LHC}.
\newblock {\em JINST}, 3:S08005, 2008.

\bibitem{vanHerwijnen:2010zz}
Eric van Herwijnen.
\newblock {The LHCb trigger}.
\newblock {\em PoS}, ICHEP2010:027, 2010.

\bibitem{Aad:2008zzm}
ATLAS Collaboration.
\newblock {The ATLAS Experiment at the CERN Large Hadron Collider}.
\newblock {\em JINST}, 3:S08003, 2008.

\bibitem{Adolphi:2008zzk}
CMS Collaboration.
\newblock {The CMS experiment at the CERN LHC}.
\newblock {\em JINST}, 3:S08004, 2008.

\bibitem{Aamodt:2008zz}
ALICE Collaboration.
\newblock {The ALICE experiment at the CERN LHC}.
\newblock {\em JINST}, 3:S08002, 2008.

\bibitem{Khachatryan:2010yr}
CMS Collaboration.
\newblock {Prompt and non-prompt $J/\psi$ production in $pp$ collisions at
  $\sqrt{s} = 7$ TeV}.
\newblock {\em Eur. Phys. J.}, C71:1575, 2011.

\bibitem{Aad:2011sp}
ATLAS Collaboration.
\newblock {Measurement of the differential cross-sections of inclusive, prompt
  and non-prompt $J/\psi$ production in proton-proton collisions at $\sqrt{s} =
  7$ TeV}.
\newblock {\em Nucl. Phys.}, B850:387--444, 2011.

\bibitem{Aaij:2011jh}
LHCb Collaboration.
\newblock {Measurement of $J/\psi$ production in $pp$ collisions at
  $\sqrt{s}=7$ TeV}.
\newblock {\em Eur. Phys. J.}, C71:1645, 2011.

\bibitem{Aamodt:2011gj}
ALICE Collaboration.
\newblock {Rapidity and transverse momentum dependence of inclusive $J/\psi$
  production in $pp$ collisions at $\sqrt{s} = 7$ TeV}.
\newblock {\em Phys. Lett.}, B704:442--455, 2011.

\bibitem{Aaij:2011yc}
LHCb Collaboration.
\newblock {Observation of $J/\psi$ pair production in $pp$ collisions at
  $\sqrt{s}=7$ TeV}.
\newblock 2011.

\bibitem{LHCb-CONF-2010-013}
LHCb Collaboration.
\newblock {Prompt charm production in $pp$ collisions at $\sqrt{s} = 7$ TeV}.
\newblock Dec 2010.

\bibitem{ATLAS-CONF-2011-017}
ATLAS Collaboration.
\newblock {Measurement of $D^{(*)}$ meson production cross sections in pp
  collisions at $\sqrt{s}=7$ TeV with the ATLAS detector}.
\newblock Technical Report ATLAS-CONF-2011-017, Geneva, Mar 2011.

\bibitem{Aaij:2011ju}
LHCb Collaboration.
\newblock {First observation of $B_s -> D_{s2}^{*+} X \mu \nu$ decays}.
\newblock {\em Phys. Lett.}, B698:14--20, 2011.

\bibitem{Bediaga:2010gn}
LHCb Collaboration.
\newblock {Measurement of $\sigma(pp \to b \bar{b} X)$ at $\sqrt{s}=7$ TeV in
  the forward region}.
\newblock {\em Phys.Lett.}, B694:209--216, 2010.

\bibitem{LHCb-CONF-2011-001}
LHCb Collaboration.
\newblock {$b$-hadron lifetime measurements with exclusive $b\to J/\psi X$
  decays reconstructed in the 2010 data}.
\newblock Mar 2011.
\newblock LHCb-CONF-2011-001.

\bibitem{LHCb-CONF-2011-036}
LHCb Collaboration.
\newblock {Studies of beauty baryons decaying to $D^{0}p\pi^{-}$ and
  $D^{0}pK^{-}$}.
\newblock Jul 2011.
\newblock LHCb-CONF-2011-036.

\bibitem{LHCb-CONF-2011-017}
LHCb Collaboration.
\newblock {Measurement of the $B_c^+$ to $B^+$ production cross-section ratios
  at $\sqrt{s}$ = 7 TeV in LHCb}.
\newblock Apr 2011.
\newblock LHCb-CONF-2011-017.

\bibitem{LHCb-CONF-2011-028}
LHCb Collaboration.
\newblock {Measurement of $b$-hadron production fractions in 7 TeV
  centre-of-mass energy $pp$ collisions}.
\newblock Jun 2011.
\newblock LHCb-CONF-2011-028.

\bibitem{Aaij:2011hi}
LHCb Collaboration.
\newblock {Determination of $f_s/f_d$ for 7 TeV $pp$ collisions and a
  measurement of the branching fraction of the decay $B_d\to D^-K^+$}.
\newblock 2011.

\bibitem{LHCb-CONF-2011-034-001}
LHCb Collaboration.
\newblock {Average $f_s/f_d$ b-hadron production fraction for 7 TeV $pp$
  collisions}.
\newblock 2011.

\bibitem{LHCb-CONF-2011-053}
LHCb Collaboration.
\newblock {Observations of Orbitally Excited $B_{(s)}^{**}$ Mesons}.
\newblock Oct 2011.
\newblock LHCb-CONF-2011-053.

\bibitem{Asner:2010qj}
Heavy Flavor~Averaging Group.
\newblock {Averages of $b$-hadron, $c$-hadron, and $\tau$-lepton Properties}.
\newblock 2010.

\bibitem{LHCb-CONF-2011-054}
LHCb Collaboration.
\newblock {Measurement of the Charm Mixing Parameter $y_{CP}$ in Two-Body Charm
  Decays}.
\newblock Aug 2011.
\newblock LHCb-CONF-2011-054.

\bibitem{LHCb-CONF-2011-046}
LHCb Collaboration.
\newblock {Measurement of the $CP$ Violation Parameter $\cal{A}_{\Gamma}$ in
  Two-Body Charm Decays}.
\newblock Jul 2011.
\newblock LHCb-CONF-2011-046.

\bibitem{LHCb-CONF-2011-061}
LHCb Collaboration.
\newblock {A search for time-integrated $CP$ violation in $D^0\to h^+ h^-$
  decays}.
\newblock Nov 2011.
\newblock LHCb-CONF-2011-061.

\bibitem{Buras:2010wr}
Andrzej~J. Buras.
\newblock {Minimal flavour violation and beyond: Towards a flavour code for
  short distance dynamics}.
\newblock {\em Acta Phys. Polon.}, B41:2487--2561, 2010.

\bibitem{Aaij:2011rja}
LHCb Collaboration.
\newblock {Search for the rare decays $B^0_{(s)} \to \mu^{+} \mu^{-}$}.
\newblock {\em Phys.Lett.}, B699:330--340, 2011.

\bibitem{LHCb-CONF-2011-037}
LHCb Collaboration.
\newblock {Search for the rare decays $B^0_{(s)} \to \mu^{+} \mu^{-}$ with 300
  pb$^{-1}$ at LHCb}.
\newblock Jul 2011.
\newblock LHCb-CONF-2011-037.

\bibitem{Chatrchyan:1371756}
CMS collaboration.
\newblock {Search for $B^{0}_{s} \to \mu^{+}\mu^{-}$ and $B^{0} \to
  \mu^{+}\mu^{-}$ decays in $pp$ collisions at $\sqrt{s} = 7$ TeV}.
\newblock Aug 2011.

\bibitem{CMS-PAS-BPH-11-019}
LHCb and CMS Collaborations.
\newblock {Search for the rare decay $B^{0}_{s}\to \mu^{+}\mu^{-}$ at the LHC
  with the CMS and LHCb experiments}.
\newblock Aug 2011.

\bibitem{Aaltonen:2011fi}
CDF Collaboration.
\newblock {Search for $B_s \to \mu^+\mu^-$ and $B_d \to \mu^+\mu-$ Decays with
  CDF II}.
\newblock 2011.

\bibitem{Akeroyd:2011kd}
A.~G. Akeroyd, F.~Mahmoudi, and D.~Martinez Santos.
\newblock {The decay $B_s\to\mu^+\mu^-$: updated SUSY constraints and
  prospects}.
\newblock 2011.

\bibitem{LHCb-CONF-2011-038}
LHCb Collaboration.
\newblock {Angular analysis of $B^0\rightarrow K^{*0}\mu^+\mu^-$}.
\newblock Aug 2011.

\bibitem{Aaij:2011ex}
LHCb Collaboration.
\newblock {Search for the lepton number violating decays $B^{+}\rightarrow
  \pi^- \mu^+ \mu^+$ and $B^{+}\rightarrow K^- \mu^+ \mu^+$}.
\newblock 2011.

\bibitem{LHCb-CONF-2011-055}
LHCb Collaboration.
\newblock {Measurement of the ratio of branching fractions $\mathcal{B}(B_d\to
  K^{*0}\gamma)/\mathcal{B}(B_s\to \phi\gamma)$ with the LHCb experiment at
  $\sqrt{s}=7$~TeV}.
\newblock Aug 2011.
\newblock LHCb-CONF-2011-055.

\bibitem{Aaij:2011fx}
LHCb Collaboration.
\newblock {First observation of $B_s \to J/\psi f_0(980)$ decays}.
\newblock {\em Phys.Lett.}, B698:115--122, 2011.

\bibitem{Li:2011pg}
Belle Collaboration.
\newblock {Observation of $B_s^0\to J/\psi f_0(980)$ and Evidence for $B_s^0\to
  J/\psi f_0(1370)$}.
\newblock {\em Phys. Rev. Lett.}, 106:121802, 2011.

\bibitem{Aaltonen:2011nk}
CDF Collaboration.
\newblock {Measurement of branching ratio and $B_s^0$ lifetime in the decay
  $B_s^0\to J/\psi f^0(980)$ at CDF}.
\newblock 2011.

\bibitem{LHCb-CONF-2011-049}
LHCb Collaboration.
\newblock {Tagged time-dependent angular analysis of $B_s\to J/\psi\phi$ decays
  with 337~pb$^{-1}$ at LHCb}.
\newblock Aug 2011.
\newblock LHCb-CONF-2011-049.

\bibitem{LHCb-CONF-2011-051}
LHCb Collaboration.
\newblock {Measurement of $\phi_s$ in $B_s \to J/\psi f_0(980)$}.
\newblock Sep 2011.
\newblock LHCb-CONF-2011-051.

\bibitem{LHCb-CONF-2011-056}
LHCb Collaboration.
\newblock {Combination of $\phi_s$ measurements from $B_{s}^{0}\to J/\psi\phi$
  and $B_{s}^{0}\to J/\psi f_{0}(980)$}.
\newblock Aug 2011.
\newblock LHCb-CONF-2011-056.

\bibitem{Bs2JpsiKs}
Kristof De~Bruyn, Robert Fleischer, and Patrick Koppenburg.
\newblock \protect{Extracting $\gamma$ and Penguin Topologies through CP
  Violation in $B_s^0\rightarrow J/\psi K_{\mathrm S}$}.
\newblock {\em Eur.~Phys.~J.~C}, 70(4):1025--1035, December 2010.
\newblock \href{http://arxiv.org/abs/1010.0089}{arXiv:1010.0089 [hep-ph]}.

\bibitem{Bs2JpsiKs_conf}
Kristof De~Bruyn, Robert Fleischer, and Patrick Koppenburg.
\newblock \protect{Extracting $\gamma$ and Penguin Parameters from
  $B_s^0\rightarrow J/\psi K_{\mathrm S}$}.
\newblock {\em Talk at CKM2010, Warwick, UK, September 6--10}, 2010.
\newblock \href{http://arxiv.org/abs/1012.0840}{arXiv:1012.0840 [hep-ph]}.

\bibitem{CDFBs}
CDF Collaboration.
\newblock {Observation of $B_s\to J/\psi K^\ast$ and $B_s\to J/\psi K_S$
  Decays}.
\newblock {\em Phys. Rev.}, D83:052012, 2011.

\bibitem{LHCb-CONF-2011-048}
LHCb Collaboration.
\newblock {Measurement of the $B^0_s \to J/\psi K^0_s$ branching fraction}.
\newblock Sep 2011.
\newblock LHCb-CONF-2011-048.

\end{thebibliography}

\end{document}